\documentclass[final,english]{bullsrsl}[2022/06/15]

\usepackage[latin1]{inputenc}
\usepackage[T1]{fontenc}

\usepackage{natbib} 
\usepackage{graphicx}
\usepackage{subfigure}

\begin{document}
\title{  On the potential of Carbon-Enhanced Metal-Poor stars for Galactic Archaeology\footnote{Invited talk} }

\author[affil={1}, corresponding]{Aruna}{Goswami}
\author[affil={1,2}]{J}{Shejeelammal}
\author[affil={1}]{Partha Pratim}{Goswami}
\author[affil={1}]{Meenakshi}{Purandardas}
\affiliation[1]{Indian Institute of Astrophysics, Koramangala, Bangalore
    560034, India}
\affiliation[2]{ Universidade de Sao Paulo, Instituto de Astronomia, 
Rua do Matao 1226, Cidade Universitaria, 05508-900, SP, Brazil   }
\correspondance{aruna@iiap.res.in}
\date{ May 3, 2023}
\maketitle


%

\begin{abstract}
The low-mass   metal-poor  stars in the Galaxy  that preserve in 
their atmosphere, the chemical imprints of the gas clouds from which they were formed  
can be used as probes to get insight into the origin and evolution of elements in the 
early galaxy, early star formation and  nucleosynthesis.  
Among the metal-poor stars, a large fraction,  the  so-called 
carbon-enhanced metal-poor (CEMP) stars exhibits high abundance of carbon. These stars  
exhibit diverse 
abundance patterns, particularly for heavy elements, based on which  they are 
classified into different groups. The  diversity of  abundance patterns  points at 
different formation scenarios. Hence, accurate 
classification of CEMP stars and knowledge of their distribution is essential 
to understand  the role and contribution of each group. While CEMP-s and CEMP-r/s stars 
can be used to get insight into binary interactions at  very low metallicity, 
CEMP-no stars 
can be used to probe the properties of the  first stars and early nucleosynthesis. 
To exploit the full potential of CEMP stars for Galactic archaeology  a  homogeneous 
analysis of each class  is extremely important. Our 
efforts towards, and contributions to providing an improved classification scheme for  
accurate classification of CEMP-s  and CEMP-r/s stars  and  in characterizing the 
companion asymptotic giant branch (AGB) stars of CH, CEMP-no, CEMP-s and CEMP-r/s binary 
systems are discussed. Some recent results obtained based on low- and high-resolution 
spectroscopic analysis of a large number of potential CH and CEMP star candidates 
are highlighted. 

\end{abstract}


\section{Introduction}
A number of large sky survey programs in the past, such as, HK 
survey \citep{Beers_1985, Beers_1992, Beers_2007, Beers_1999}, 
Hamburg/ESO Survey (HES;  \citealt{Christlieb_2001a, Christlieb_2001b, Christlieb_2003, 
Christlieb_2008}),
Sloan Digital Sky Survey (SDSS; \citealt{York_2000}),  LAMOST
(Large Sky Area Multi-Object Fiber Spectroscopic Telescope) survey
\citep{Cui_2012, Deng_2012,  Zhao_2012} were dedicated for identifying the most 
metal-poor stars. These surveys revealed  that  a significant fraction ($\sim$20\% of 
Very Metal-Poor (VMP), [Fe/H]$<$$-$2) stars in the Galaxy are CEMP stars 
\citep{Rossi_1999, Christlieb_2003, Lucatello_et_al_2005, Lucatello_et_al_2006, Carollo_2012}
and that the fraction of these stars  increases with decreasing metallicity
$\sim$40\% for [Fe/H]$<$$-$3 and $\sim$75\% for [Fe/H]$<$$-$4
\citep{Yong_2013b, Lee_2013, Aoki_2013, placco_et_al_2014_cfe, frebel_norris_2015}.

\citet{beers2005discovery} first introduced  four sub-classes of CEMP stars considering the 
abundances of carbon and  two {\it{n}}-capture elements, Ba (a representative s-process element) and Eu 
(a representative r-process element). Subsequently, various authors \citep{Jonsell_et_al_2006, Masseron_et_al_2010, Norris_et_al_2010, Bisterzo_2011, Bonifacio_2015, Maeder_2015b, Yoon_et_al_2016,  Abate_et_al_2016, Hansen_2016a, Frebel_review_2018, hansen_et_al_2019, Skuladottir_et_al.2020, Goswami_et_al_1_2021} put forward different criteria for the identification and classification of the CEMP stars.  The four primary sub-classes that are  based on the level of relative enrichment of neutron-capture elements are:  CEMP-s (shows enhanced 
abundances of s-process elements), CEMP-r (shows strong enhancement of r-process elements), 
CEMP-r/s (shows enhancement of both s- and r-process elements) and CEMP-no (does not show 
any enhanced abundance of neutron-capture elements). The  diverse abundance patterns  
point at  different formation scenarios for the respective sub-classes. 
The  fractions and properties of CEMP sub-classes  provide a unique opportunity to 
probe the 
formation and evolution of the Galactic  halo  and its building blocks.  
For this reason, the CEMP stars alongside the  halo tracers   CH stars  
 have been  extensively studied in the last few decades. 

\par CH stars ($-$2$\leq$[Fe/H]$\leq$$-$0.2) are characterized by  strong CH and C$_{2}$ molecular bands, C/O$>$1 and strong features due to the neutron-capture elements \citep{Keenan_1942}. These are  mostly high velocity objects (V$_{r}$ > 100 km s$^{-1}$),  and members of the  Galactic halo. 
A few CH stars  are also found in Globular clusters.  Sub-giant CH stars have moderate velocities and they exhibit  old disk behaviour rather than the halo. These objects are believed to be the progenitors of metal-deficient Ba stars.   Long term radial velocity study by \citet{McClure_woodsworth_1990}  
showed that most of the CH stars are binaries with companions that are now presumably white dwarfs. 
The heavy s-process elements in giant CH stars are found to be more enhanced than the 
light s-process elements \citep{Karinkuzhi_goswami_2014, Karinkuzhi_goswami_2015}, 
in agreement with theoretical modelling of s-process in stars and its dependence on the metallicity
\citep{Busso_et_al_1999, Busso_et_al_2001, Cristallo_et_al_2016}.

The CEMP-s stars are the metal-poor analogue of Ba and CH stars 
\citep{Lucatello_et_al_2005, Abate_et_al_2016}.  A number of  long-term  radial velocity 
monitoring studies have shown that most of the CEMP-s stars are binaries 
\citep{Lucatello_et_al_2005, Starkenburg_2014, jorissen2016rv, Hansen_2016a}, 
supporting the pollution from companion AGB stars. Comparisons of  observed abundances 
in the CEMP-s stars with theoretical  model predictions   confirm the binary 
mass-transfer from the AGB companion \citep{Bisterzo_2011, Placco_2013, Hollek_2015, 
Shejeelmmal_et_al_2020, Shejeelammal_goswami_2021, Shejeelammal_goswami_2022, 
Goswami_et_al_1_2021, goswami_goswami_ana_2022}.

\par  The  CEMP-r/s stars  show abundance enhancement in both slow (s-) and 
rapid (r)-neutron  capture process elements.   
Since most of the CEMP-r/s stars are also found to be binaries 
just like CEMP-s stars, binary mass transfer from the AGB companion is thought to be 
the reason 
for their origin as well, however the presence of r-process component posed a 
challenge \citep{Jonsell_et_al_2006, Herwig_et_al_2011, Abate_et_al_2016}. It was proposed 
by \citet{Cowan_1977} that  the intermediate neutron-capture process the so-called 
i-process, with neutron density between  s- and r- process can produce both s- and 
r- process  elements at a single stellar site. Many studies  have
successfully used  model yields of i-process in low-mass low-metallicity AGB stars 
to account 
for the observed abundance patterns in CEMP-r/s stars \citep{Hampel_et_al_2016, Hampel_et_al_2019, Goswami_et_al_1_2021, Shejeelmmal_et_al_2020, Shejeelammal_goswami_2021,   Shejeelammal_goswami_2022}. Although there are  several suggestions for the i-process sites, such as, 
super-massive AGB stars \citep{Doherty_et_al_2015, Jones_2016}, low-mass low-metallicity stars
\citep{Campbell_2008, Campbell_2010, Cruz_2013, Cristallo_et_al_2016}, and 
rapidly accreting white dwarfs \citep{Herwig_2014, Denissenkov_et_al_2017} etc., the exact
astrophysical site for the i-process is not yet confirmed \citep{Frebel_review_2018, Koch_et_al_2019}.

\par  CEMP-r stars are extremely rare. In spite of several studies on  the formation 
of CEMP-r stars, the origin of these stars are  not yet clearly understood; 
observations of more 
such stars are required to constrain their  exact origin.

 The CEMP-no stars with enhanced carbon abundance and  no signatures of heavy 
neutron-capture  elements, are believed to be the most  chemically primitive  objects 
presently known \citep{Norris_2019, Yoon_2020}. These stars occupy the lowest 
metallicity tail ([Fe/H]$<$$-$3) of the Metallicity Distribution Function (MDF) of 
the Galaxy \citep{Yong_2013, Bennassuti_2017, Yoon_2019, Norris_2019}. 
Among the fourteen known CEMP stars with [Fe/H]$<$$-$4.5, twelve stars belong to the 
CEMP-no group. 
Being the oldest stars that can be directly observed at present, 
valuable information can be derived from them, such as the formation and nucleosynthesis 
pathways of the first stars,  initial conditions for star and  galaxy
formation \citep{klessen_et_al_2012, frebel_norris_2015, Bonifacio_2015, 
Hansen_2016a, Yoon_2020, Yoon_2018},
large galaxy assembly (e.g. \citet{Carollo_2007}; Beers et al. (2012)), and early 
nucleosynthesis (\citet{hansen_et_al_2011}; \citet{gil_pons_p_et_al_2021}; \cite{qian_2022}).
Further, these stars are 
expected to shed light on the physical processes  resulting in the transitions from 
massive Population III (Pop III) stars to normal Population II (Pop II) stars \citep{Salvadori_2007, Hartwig_2015, Bennassuti_2017}.

Different scenarios have been suggested by various authors for the origin of CEMP-no 
stars (\citealt{Hansen_2015}, \citealt{Yoon_et_al_2016} and references there in), 
however, the exact  origin of these stars  still remains unknown. According to some 
authors, the CEMP-no stars are generally  single stars, suggesting an intrinsic 
origin, i.e, formed  from a pre-enriched natal cloud  of ISM \citep{Spite_2013, 
Starkenburg_2014, Bonifacio_2015, Hansen_2016a, Yoon_et_al_2016}.  Suggested progenitors 
for the CEMP-no stars also include: 
(i) faint supernovae that underwent mixing and fall back. 
 \citep{Umeda_2003, Umeda_2005, Tominaga_2014},
(ii) extremely metal-poor fast rotating massive stars or spinstars that enriched the ISM 
with C, N, and O  through strong stellar winds \citep{Meynet_2006, Frischknecht_2012, 
Maeder_2015a, Choplin_2017}, and 
(iii) i-process in massive Pop III stars \citep{Clarkson_2018}. 
A recent study by \cite{Arentsen_2019b} have shown that the binary mass-transfer from an 
extremely metal-poor AGB companion could also be a possible progenitor of the CEMP-no stars.
If the binary-mass-transfer scenario holds true for the origin of CEMP-no stars, this will 
help to understand extremely metal-poor AGB stars and provide robust observational 
constraints to the AGB nucleosynthesis and mass-transfer models for metallicity [Fe/H]$<$$-$2. 

Since most of the metal-poor stars in the halo trace their origin to the dwarf satellite galaxies \citep{frebel_norris_2015, Frebel_2023},  they are ideal tools to study the Galactic halo assembly. 
Based on the kinematic, chemical, and dynamical properties of a sample of 644 CEMP stars, \cite{Zepeda_2023} have shown that the CEMP-no stars are not in-situ to the Galaxy, but were originally born in satellite galaxies that were  accreted and disrupted by the Milky Way. 
This is also in support of  the findings of \cite{Yoon_2019}.  Several authors have discussed 
the relevance and scope of enhancing medium- and high-resolution spectroscopic studies of 
different sub-groups of CEMP stars to further our understanding of early Galactic chemical evolution and Galactic halo formation (eg. \citealt{Hansen_2015,  Norris_2019, 
Yoon_2019, Yoon_2020, Zepeda_2023}). 
Along this line, we had conducted  high-resolution spectroscopic studies of several
 potential CH and CEMP star candidates identified from low-resolution spectroscopic 
studies with the aim of deriving clues to their origin. 
   
 In section 2, we have discussed the limitations  in classifying CEMP stars based on low-resolution spectra. Shortcomings of current classification criteria for CEMP-s and CEMP-r/s stars  is 
 discussed in section 3. A revised classification scheme for identifying the CEMP-s and 
 CEMP-r/s stars is also presented in this section. In section 4, we have discussed about 
 charaterizing the companion AGBs of the CH, CEMP-s and CEMP-r/s stars. A discussion on deriving 
 clues for the origin of CEMP-r/s stars from abundance patterns is presented in section 5. Observational evidence for multiple origins for CEMP-no stars is discussed in section 6. 
 The unique star 
 HE 1005-1439 that shows evidence of occurrence of dual nuclesynthesis is discussed in 
 section 7. A few critical  issues in  analysis of CEMP stars are discussed in section 8. 
 The conclusions are presented in section 9.
             
\section{Limitations in classifying  CEMP stars using low-resolution spectra}
Due to the intrinsic  weakness of the absorption features, using low-resolution samples, 
it is difficult to  distinguish between different sub-classes of CEMP stars.  
This is illustrated in figure 1, where a few examples of low-resolution (R $\sim$ 1300) 
spectra  of CEMP stars obtained with HCT/HFOSC  are shown. Detail analysis  of these 
spectra 
can be found in \citet{Goswami2005CH, Goswami2010CH}. Visual inspection shows 
that the spectra look alike, and hence, difficult to identify if they belong to 
different sub-classes. The spectra  are characterized by strong features of carbon 
molecules, and lines due to Ba and Sr are only marginally detectable.  However, high 
resolution spectroscopic analysis have shown that HE 0319-0215 and HE 1305+0132 are 
CEMP-s star and 
HE 0017+0058 is a CEMP-r/s star  \citep{Goswami_et_al_1_2021}. Although the origin of 
both these classes  can be explained based on a binary picture with a now invisible white 
dwarf companion, the progenitors AGBs have very different nucleosynthetic and evolution history. 

\begin{figure}
    \centering
    \includegraphics[scale=0.40]
    {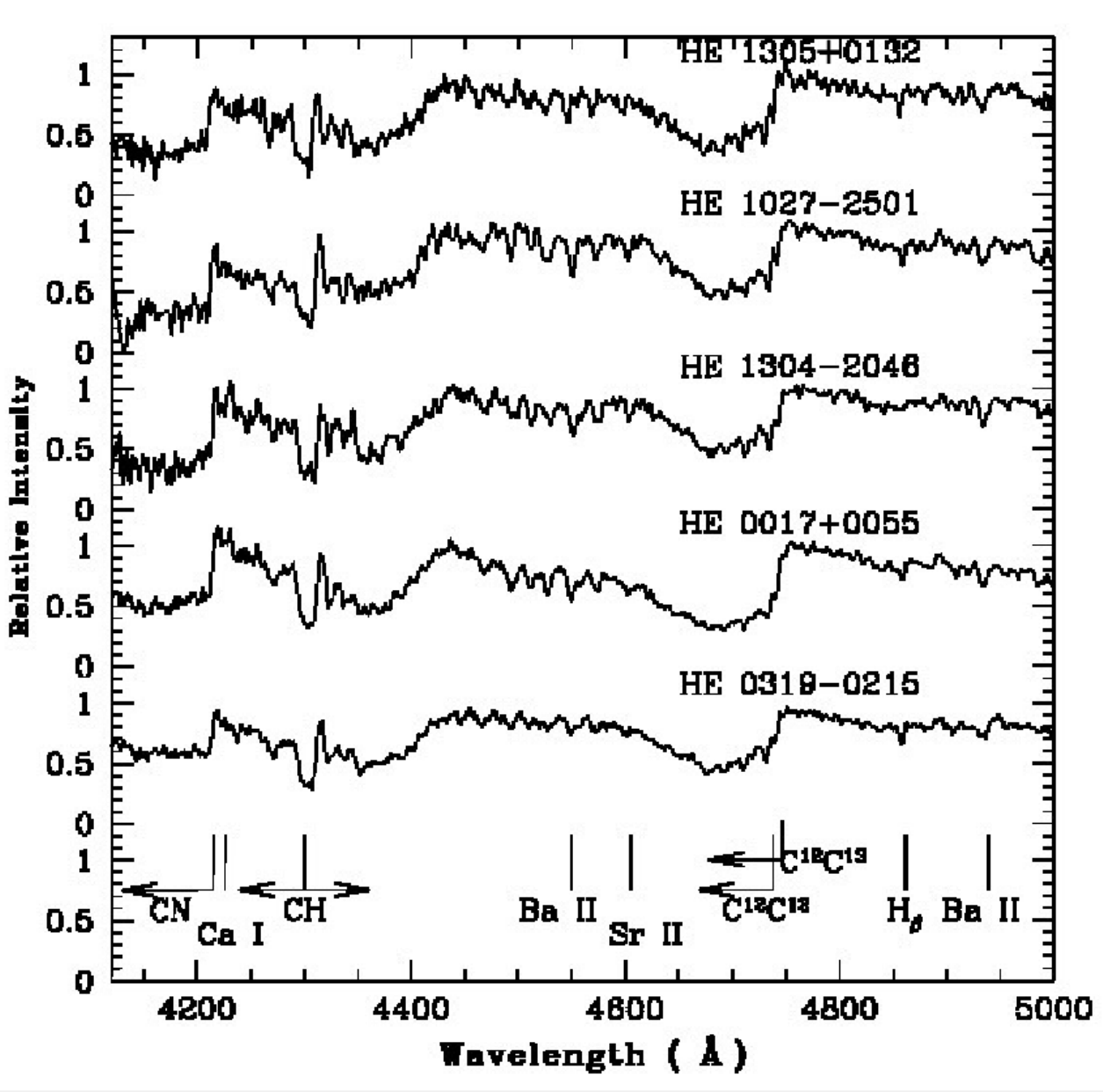}
    \bigskip
    \begin{minipage}{12cm}
    \caption{ Low resolution (R $\sim$ 1300) spectra of a few CEMP stars. The spectra 
characterized by strong  carbon molecular features  look alike  and  difficult to classify them into 
different CEMP sub-classes  based on the low-resolution spectra. 
Follow-up  high resolution spectroscopic 
studies have confirmed the objects HE~ 0319$-$0215 and HE~1305+0132 as CEMP-s stars and 
HE~0017+0058  as a CEMP-r/s star. }
 \label{fig:1}
 \end{minipage}
 \end{figure}

\section{Classification of CEMP stars: shortcomings of the current classification criteria 
and a revised classification scheme  for identifying CEMP-s and CEMP-r/s stars}

Among the CEMP stars,  distinguishing the CEMP-{\it{s}} and CEMP-{\it{r/s}} stars is 
crucial in order to understand the various astrophysical and nucleosynthetic processes 
responsible for the abundance patterns of the two sub-classes. The object  HD~145777  
in our sample, particularly
posed a challenge in this regard. This object could be classified as a CEMP-r/s star
 based on the  classification schemes of \citet{beers2005discovery}, 
\citet{Frebel_review_2018}, 
and  \citet{hansen_et_al_2019}, and  as a CEMP-{\it{s}} star if the classification 
criteria   of \citet{Abate_et_al_2016} are adopted. Based on a parametric model based 
analysis  we have found  that in order to fit the observed abundances, a model with 
a neutron density of 
10$^{10}$ cm$^{-3}$ is required \citep{Goswami_et_al_1_2021} which  is towards the 
higher limit of the neutron density required for {\it{s}}-process nucleosynthesis. To 
remove the uncertainties and  confusion regarding its classification, we have revisited 
the classification schemes of CEMP-{\it{s}} and CEMP-{\it{r/s}} stars found in the 
literature. From a detailed investigation  we have seen that none of the existing 
classification criteria could clearly  distinguish the CEMP-{\it{s}} and CEMP-{\it{r/s}}
 stars.   
We have examined if   abundance ratios such as [hs/ls] (where hs implies heavy s-process 
elements and ls the light s-process elements), [Sr/Ba], [Ba/Eu], [La/Eu], [La/Ce],  
[As/Ge] and [Se/Ge] could be used as classifiers to distinguish the CEMP-s and 
CEMP-r/s stars effectively. 
We note that the [hs/ls] peaks at different values giving higher values for 
CEMP-{\it{r/s}} stars than CEMP-{\it{s}} stars. However, the CEMP-{\it{s}} and 
the CEMP-{\it{r/s}} stars are also found to show an overlap in the range 
0.0~$<$~[hs/ls]~$<$1.5.  Hence,  this ratio alone cannot
be used as  a definitive classifier of CEMP-{\it{s}} and CEMP-{\it{r/s}} stars.   
Based on our investigation, we suggest that an object satisfying
the criteria 0.0~$<$~[Ba/Eu]~$<$~1.0 and 0.0~$<$~[La/Eu]~$<$~0.5 can be
classified as a CEMP-{\it{r/s}} star; however, if  [La/Eu] = 0.6 $\pm$ 0.1
the condition  [Eu/Fe]~$>$~1.0 also needs to be satisfied for the star to be a
CEMP-{\it{r/s}} star. We proposed, the  best criteria to distinguish the CEMP-{\it{s}}
 and CEMP-{\it{r/s}} stars as:
\begin{itemize}
      \item CEMP: [C/Fe] $\geq$ 0.7
\vskip 0.1cm
      \item CEMP-{\it{r/s}}: [Ba/Fe] $\geq$ 1.0, [Eu/Fe] $\geq$ 1.0
\vskip 0.1cm
      \begin{itemize}
            \item [i)] 0.0~$\leq$~[Ba/Eu]~$\leq$~1.0 and/or 0.0~$\leq$~[La/Eu]~$\leq$~0.7;
      \end{itemize}
\vskip 0.1cm
      \item CEMP-{\it{s}}: [Ba/Fe]~$\geq$~1.0
\vskip 0.1cm
      \begin{itemize}
            \item [i.)] [Eu/Fe]~$<$~1.0, [Ba/Eu]~$>$~0.0 and/or  [La/Eu]~$>$~0.5;
            \item [ii.)] [Eu/Fe] $\geq$ 1.0, [Ba/Eu] $>$ 1.0 and/or  [La/Eu] $>$ 0.7.
      \end{itemize}
\end{itemize}
This scheme of classification  based on the abundance ratios of three key neutron-capture
elements barium, lanthanum and europium have proven to be highly effective 
in distinguishing  the CEMP-s 
and CEMP-r/s stars. A critical investigation  on the CEMP stars formation scenarios and a thorough
discussion  on the CEMP stars classification criteria along with  the new classification scheme put 
forward by us are  presented and discussed at length in \citet{Goswami_et_al_1_2021}. 

\section{Characterizing the companion AGB stars of CH, CEMP-{\it{s}} and CEMP-{\it{r/s}} stars}
We have investigated the mass of companion AGB stars using several diagnostics such as C, N, O, Na, Mg
abundances, [hs/ls] ratio, and [Rb/Zr] ratio. The carbon isotopic ratio $^{12}$C/$^{13}$C an important 
mixing indicator is also measured  whenever possible.  Na and  Mg enrichment is 
expected to observe  if the s-process over abundance is resulting from the neutrons 
produced during the convective thermal pulses through the  
reaction $^{22}$Ne($\alpha$,n)$^{25}${Mg}.  
Several abundance ratios of certain key elements are also used as
diagnostics to understand the characteristic properties of the companion AGB stars as 
discussed in \citet{Shejeelammal_2021a, Shejeelammal_goswami_2021, 
Shejeelammal_goswami_2022, Purandardas_2021b, Goswami_et_al_1_2021}.

\subsection{The [hs/ls] ratio as an indicator of neutron source}
The [hs/ls] ratio is a useful indicator of neutron source in the 
former companion AGB star. At higher neutron exposures, more second peak (hs) s-process
elements are produced over the first peak (ls), resulting in high [hs/ls] ratio. 
As the neutron exposure increases with decreasing metallicity, [hs/ls] ratio  increases 
with decreasing metallicity, showing  anti-correlation  with metallicity. 
The anti-correlation 
of [hs/ls] suggest the operation of  $^{13}$C($\alpha$, n)$^{16}$O neutron source,  
since  $^{13}$C($\alpha$, n)$^{16}$O is found to be anti-correlated with  metallicity 
\citep{Clayton_1988, Wallerstein_1997}. Hence positive values for [hs/ls] ratio could be 
seen in low-mass, low metallicity AGB stars \citep{Goriely_2000, Busso_et_al_2001}; 
whereas, in the case of massive AGB stars (5 - 8M$_{\odot}$), where the neutron source 
is $^{22}$Ne($\alpha$, n)$^{25}$Mg, negative values for this ratio is observed 
\citep{Goriely_2005, Karakas_2010, Karakas_2012, vanRaai_2012, Karakas_2014}. 
The $^{22}$Ne($\alpha$, n)$^{25}$Mg source has smaller neutron exposure compared to 
the $^{13}$C($\alpha$, n)$^{16}$O source. Hence, in the stars where 
$^{22}$Ne($\alpha$, n)$^{25}$Mg 
operates, we expect a lower [hs/ls] ratio. The lower neutron exposure of the 
neutrons produced from the $^{22}$Ne source together with the  predictions of low 
[hs/ls] ratio in massive AGB star models  have been taken as the evidence of 
operation of  $^{22}$Ne($\alpha$, n)$^{25}$Mg in massive AGB stars. A Mg  enrichment 
is expected in the 
stars where this reaction takes place. None of the CEMP stars studied by us in 
\citet{Shejeelammal_2021a, Shejeelammal_goswami_2021, Shejeelammal_goswami_2022} shows 
such an enrichment, ruling out  the possibility of $^{22}$Ne($\alpha$, n)$^{25}$Mg 
reaction 
as a possible neutron source with respect to [hs/ls] ratio. 
All the  CEMP-r/s stars and CEMP-s stars (except HE~0920$-$0506 and HE~1354$-$2257), 
studied in the above mentioned studies  show positive values of [hs/ls] indicating 
low-mass of the companion AGB stars.

\subsection{Rb as a probe of neutron density at the s-process site: the [Rb/Zr] ratio} 
In addition to the [hs/ls] ratio, the abundance of rubidium can also provide clues to 
the mass of the companion  AGB stars. The AGB star models predict higher Rb abundances 
for massive  AGB stars  where the neutron source is $^{22}$Ne($\alpha$,n)$^{25}${Mg} 
reaction \citep{Abia_2001, vanRaai_2012}.  
Detailed nucleosynthesis models for the stars with masses between 5 - 9 M$_{\odot}$ at 
solar metallicity  predict [Rb/(Sr,Zr)]$>$0 \citep{Karakas_2012}.
A positive value of [Rb/Sr] or [Rb/Zr] ratio  indicates a higher neutron  density, 
whereas a negative value  indicates a low neutron density with  
$^{13}$C($\alpha$, n)$^{16}$O  reaction acting  as the neutron source.
A comprehensive discussion on Rb as a probe of the mass of the companion AGB star 
can be found in \citet{Shejeelammal_2021a, Shejeelammal_goswami_2021}. In these 
studies the authours have found negative values for  
[Rb/Zr] and [Rb/Sr] ratios   for  stars for 
which  these ratios could be estimated. The observed [Rb/Fe] and [Zr/Fe] ratios 
and the abundances of Rb and Zr are found to be consistent with the range normally 
observed in the  low-mass AGB stars, that indicate low-mass nature of the companion 
AGB stars.

\subsection{Determination of the mass of the companion AGB stars from  a parametric model based analysis using FRUITY}

Detail procedure of the  parametric model
based analysis to derive the mass of the companion AGB stars have been discussed  
at length in  several of our  papers  \citep{Shejeelammal_goswami_2021, 
Shejeelammal_goswami_2022, Goswami_et_al_1_2021, Purandardas_2021b, 
Goswami_&_Goswami_2023}.  
The mass of the AGB stars responsible for the observed abundances of the  stars reported 
in these papers are  
derived by  comparing the observed abundance with the predicted abundance from FRUITY
models \citep{Cristallo_2009, Cristallo_et_al_2011, Cristallo_et_al_2015} for the 
heavy elements  (Sr, Y, Zr, Ba, La, Ce, Pr, Nd, Sm and Eu).
 The best fitting mass of the companion AGB is obtained by fitting the observed 
abundance with  the parametric model function of \cite{Husti_2009} by minimizing 
the $\chi^{2}$ value.  All the  stars in these studies are found  to 
have low-mass AGB companions with M $\leq$ 3 M$_{\odot}$.

\section{Origin of CEMP-r/s stars: clues from the abundance pattern and  parametric model based analysis} 
A comprehensive discussion on         
 the parametric model based analysis and the results obtained for  the 
stellar sample studied by us 
 can be found in \citep{Goswami_et_al_1_2021, Shejeelammal_goswami_2021,  
Shejeelammal_goswami_2022}.  
The overlap of [hs/ls] ratios between the CEMP-s and CEMP-r/s stars 
\citep{Goswami_et_al_1_2021} indicates that the origin of both the groups  
owe a common astrophysical site and  process  but likely under different conditions. 
The higher [hs/ls] ratio of the  CEMP-r/s stars 
compared to the CEMP-s stars indicate a higher neutron exposure for their 
formation than the classical s-process. The tight correlation between the 
observed [Eu/Fe] and [hs/Fe] ratios in CEMP-r/s stars \citep{Goswami_et_al_1_2021}
points towards a single stellar site where both the s- and r- process elements 
could be produced simultaneously.
A higher neutron densities than that required for the classical s-process, of the 
order of  10$^{15}$ - 10$^{17}$ cm$^{-3}$, intermediate between the s- and r-process 
neutron densities, 
could be achieved when a substantial amount of hydrogen rich-material is mixed into the 
intershell region of the evolved red-giant stars (Proton Ingestion Episode, PIE) 
undergoing helium shell flash \citep{Cowan_1977}. There are several sites proposed 
to host the PIEs in order for the i-process nuclesynthesis  to take place, however 
the astrophysical  site for occurrence of i-process still remains a topic of debate.  

\par We have used the i-process model yields  [X/Fe], from \cite{Hampel_et_al_2016} 
and compared with the 
observed [X/Fe] of the CEMP-r/s stars in our sample for the neutron densities 
ranging from 10$^{9}$ to 10$^{15}$ n cm$^{-3}$.
We have considered only the neutron-capture elements for the comparison.    
The neutron-density responsible for the observed abundances in the star
is derived by fitting the observed abundance with the dilution factor 
incorporated parametric model function: 
\begin{center}
  \noindent X = X$_{i}$ . (1-d) + X$_{\odot}$ . d  
\end{center}
\noindent where X represents the final abundance, X$_{i}$ the i-process abundance, 
d is the dilution factor and
X$_{\odot}$ is the solar-scaled abundance. Here d is a free parameter that can be 
varied to find the best fit
between the model and the observational data for each constant neutron density. 
The best fit  model is found using $\chi^{2}$$\rm_{min}$ method.
Some examples of the best fits obtained for the CEMP-r/s stars are shown in Figure 2.

\begin{figure}
\centering
        \subfigure[]{%
            \label{fig:1a}
            \includegraphics[height=6.5cm,width=8.0cm]{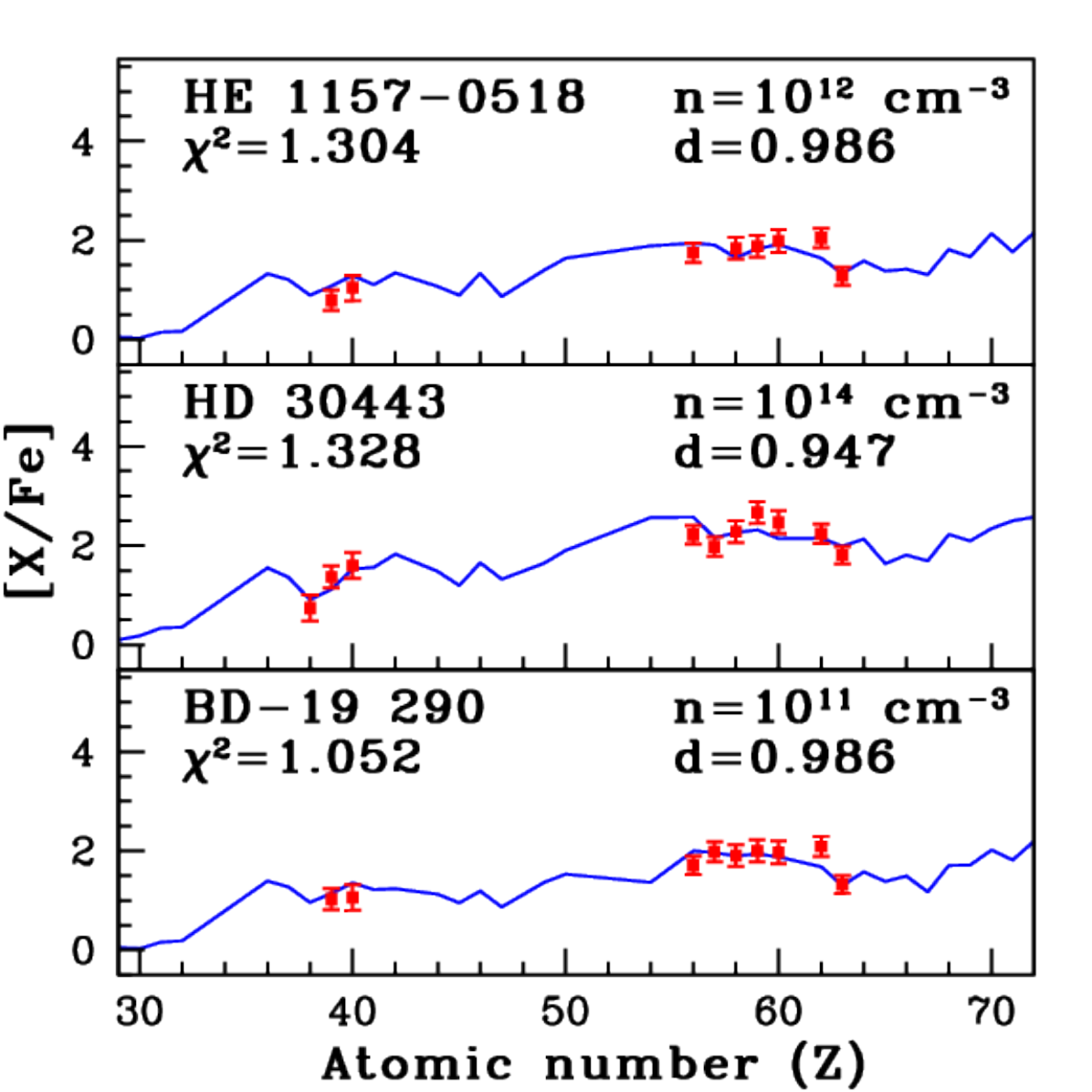}
        }%
        \subfigure[]{%
            \label{fig:1b}
            \includegraphics[height=6.5cm,width=8.0cm]{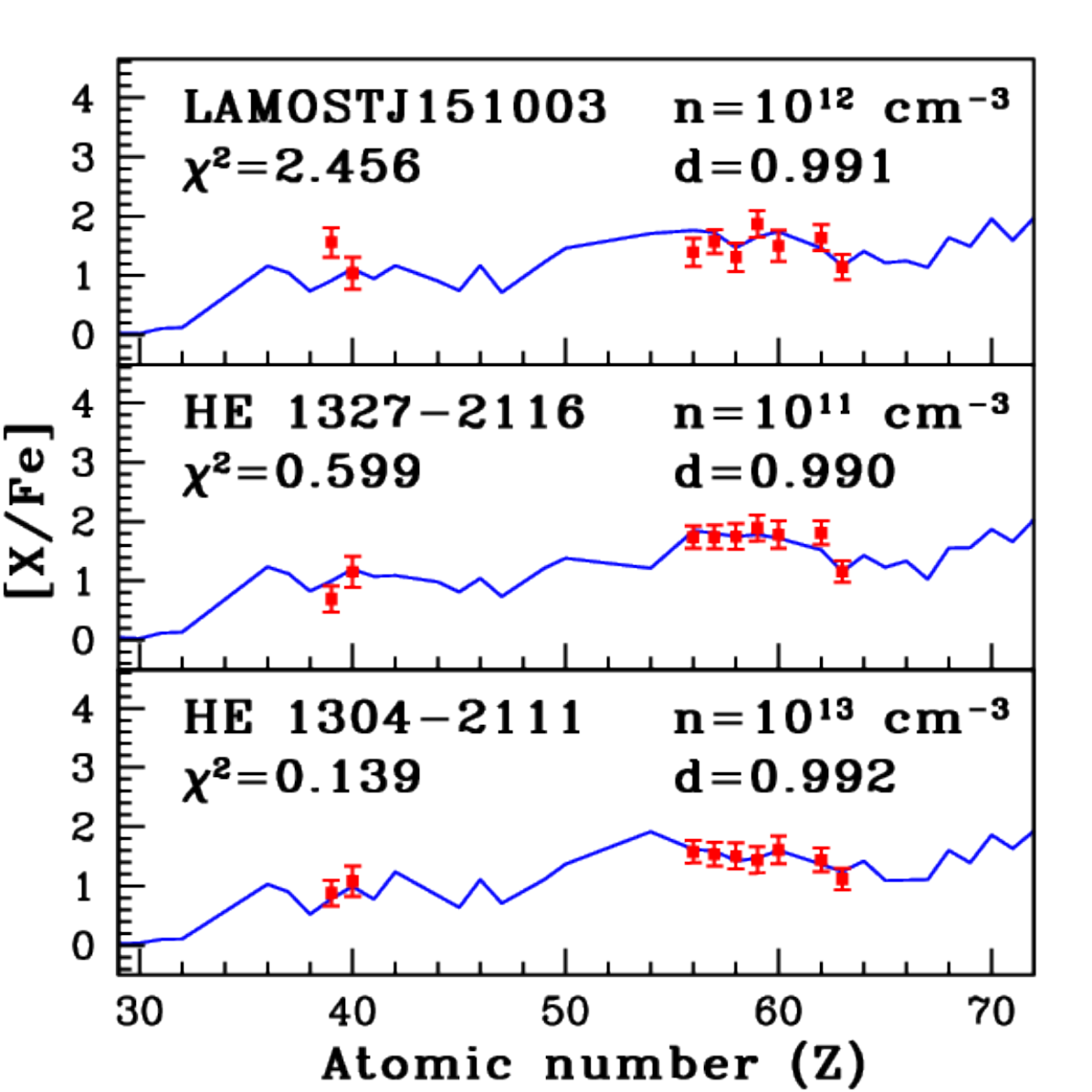}
        }\\ 
        \bigskip
        \begin{minipage}{12cm}

    \caption{Parametric model fits for CEMP-r/s stars. 
Solid curves represent the best fit for the parametric model function.
The points with error bars indicate the observed abundances.}
\label{fig:3}
\end{minipage}
\end{figure}

\section{Observational evidence for multiple origins for CEMP-no stars}
The differences in the morphology of CEMP-no stars on the Yoon - Beers diagram
(A(C) - [Fe/H] diagram),
leading to  two distinct groups, Group II and Group III CEMP-no stars  suggest
multiple formation pathways for CEMP-no stars \citep{Yoon_et_al_2016}.
The two CEMP-no stars HE~2148$-$2039 and HE~2155$-$2043  studied by us  
 \citep{Purandardas_Goswami_2021} are also found to provide observational evidence 
for multiple origins for CEMP-no stars. These two objects are found to  have 
similar metallicity ([Fe/H] of $-$ 3.32 and $-$3.46 respectively)  and enhanced 
carbon with [C/Fe] values 1.27  and 2.05 respectively. 
While Mg is found to be under abundant in HE~2148$-$2039 with
[Mg/Fe] = $-$0.35, it  is highly enhanced in HE~2155$-$2043 with [Mg/Fe] =  1.57, 
indicating different origins  as far as the ${\alpha}$-elements are concerned.
As far as the heavy elements are concerned, 
while HE~2148$-$2039 shows a very strong feature of Ba II 4934.06 \AA\, and a weak 
feature of Sr II 4215.52 \AA\,, in contrast,  the object 
HE~2155$-$2043 shows a strong feature of   Sr II 4215.52 \AA\,  and a weak feature of  
Ba II 4934.06 \AA\ (Figure 3). The estimated [Sr/Fe] and [Ba/Fe] are ($-$2.02, $-$0.84) 
for  HE~2148$-$2039 
and ($-$0.04, $-$1.64) for HE~2155$-$2043. These differences clearly indicate  
different origins for these two objects. The locations of these two stars in the 
absolute carbon abundance, A(C) versus [Fe/H] diagram, show that HE~2148$-$2039 is 
a CEMP-no Group II object and HE~2155$-$2043 is a CEMP-no Group III object
\citep{Purandardas_Goswami_2021}. 

\begin{figure}
\centering
        \subfigure[]{%
            \label{fig:1a}
            \includegraphics[height=6.5cm,width=8.0cm]{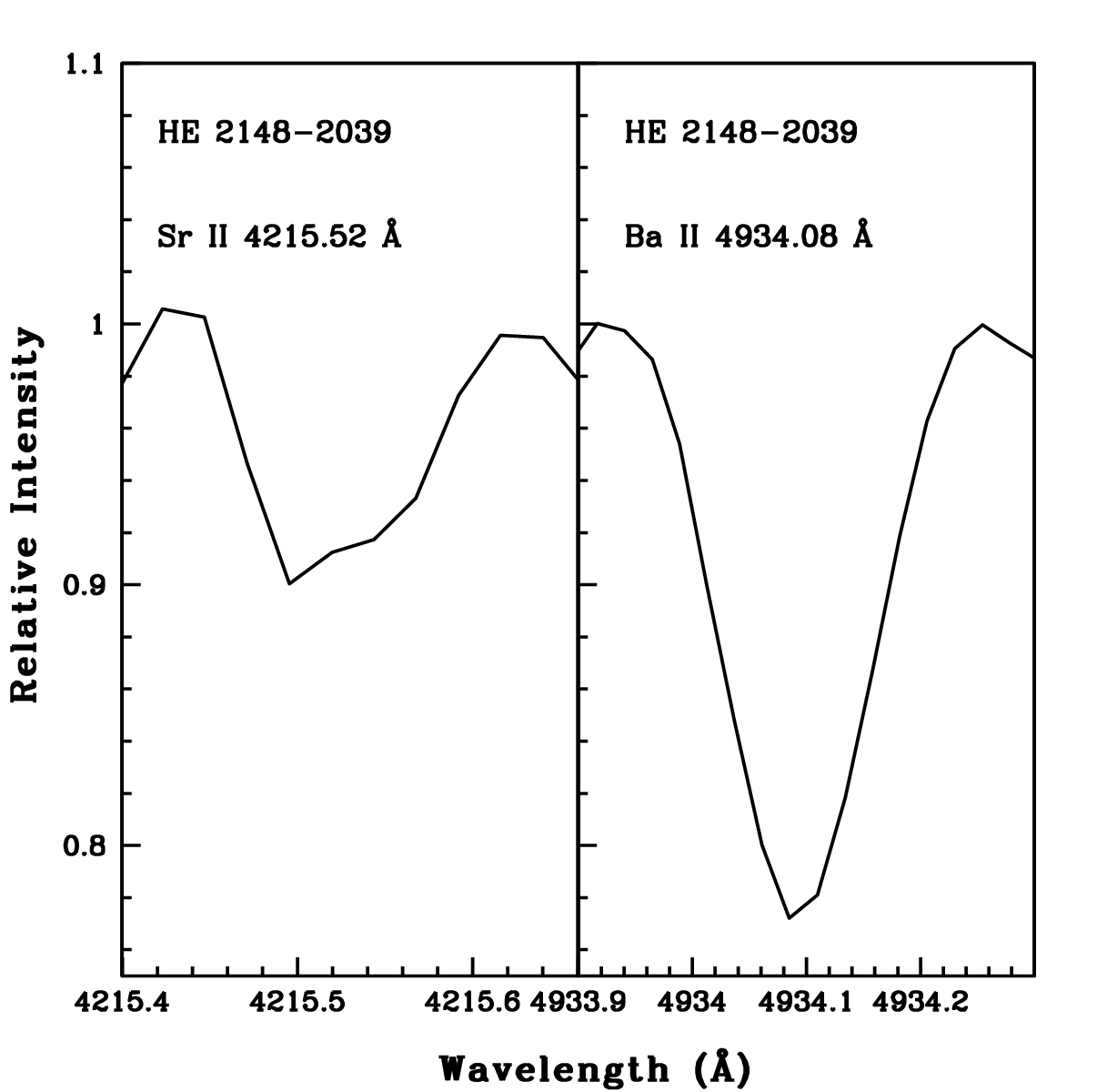}
        }%
        \subfigure[]{%
            \label{fig:1b}
            \includegraphics[height=6.5cm,width=8.0cm]{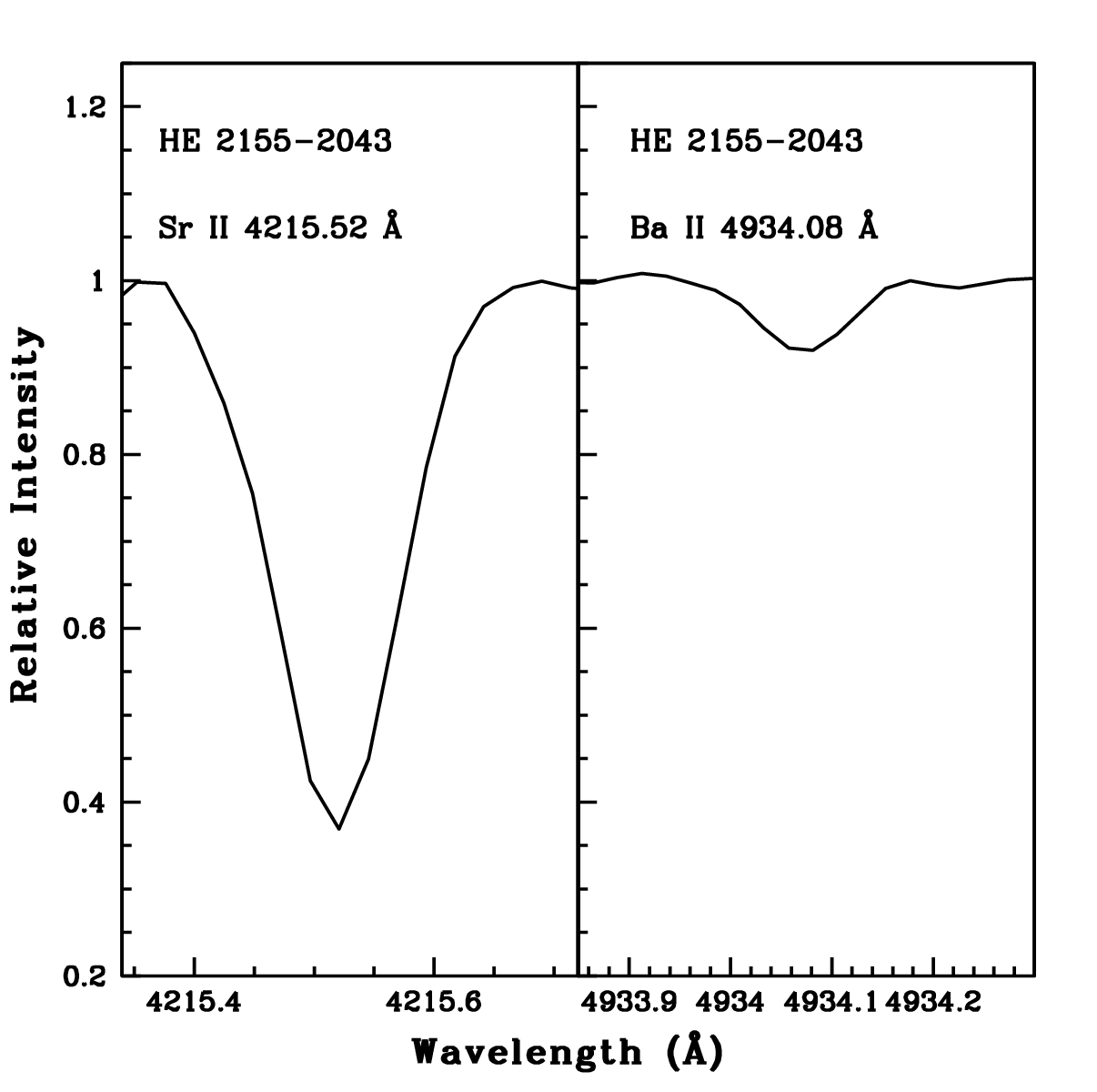}
        }\\ 
\begin{minipage}{12cm}
    \caption{ While Ba II 4934.08 \AA\,  is marginally  detectable  in HE~2155$-$2043, this feature 
    appears strong in HE~2048$-$2039. On the other hand,  Sr II 4215.52 \AA\,  feature is strong in HE~2155$-$2043 and weak in HE~2148$-$2039.   Both the objects are of similar metallicity [Fe/H] $<$ $-$3.3 and [C/Fe] $>$ 1.0. 
 }

\label{fig:4}
\end{minipage}
\end{figure}

 \section{HE~1005-1439: A unique star with dual nucleosynthesis signatures}
 
This object with a  unique abundance pattern  that shows dual nucleosynthesis 
signatures   was first reported in 
\citet{goswami_goswami_ana_2022}. The star was found to be  extremely metal-poor 
with [Fe/H] = $-$ 3.03 and  heavily enriched with 
neutron-capture elements. The observed  elemental abundances of the object could 
not be explained 
based on theoretical s-, r-, or i-process model predictions alone.  We have performed  a
parametric-model-based analysis 
of the abundances of the heavy elements which  indicated that  the star's surface 
chemical composition is being influenced by similar contributions from both 
the s- and i-processes. This forms  a new class of object of distinct abundance 
pattern with dual nucleosynthesis signatures.

The estimated radial velocities of the object from several epochs showed variations 
indicating the presence of a binary companion. Based on this observation, we  
proposed a formation scenario for HE 1005-1439 involving mass transfer from a now 
extinct AGB companion where both i- and s-process nucleosynthesis took place during 
various stages of the AGB evolution with proton ingestion episodes (PIEs) triggering 
i-process followed by s-process AGB nucleosynthesis with a few 
third-dredge-up episodes. 
A detailed study  on this object and the significance  of the results obtained are 
 presented  in \citet{goswami_goswami_ana_2022}.

\section{A few critical issues in  analysis of CEMP stars}
To understand the role of CEMP stars on Galactic chemical evolution, accurate knowledge 
of the CEMP fractions among the metal-poor stars of the Galactic halo is vital. 
The fraction is found to have different values based on the adopted definition of CEMP 
stars, i.e., either [C/Fe] $>$ 1.0 (Beers \& Chrislieb 2005), or 
[C/Fe] ${\geq}$ 0.7 \citet{aoki2007carbon}. 
A compilation of published CEMP fractions and samples of Galactic halo
stars from the past 25 years, revealed that they are not all consistent with
each other (Arentsen et al. 2022). 
Significant differences were noticed between various surveys when comparing their trends 
of [Fe/H] versus [C/Fe] and their distributions of CEMP stars. These discrepancies were 
primarily attributed to the differences in the analysis procedures and criteria 
adopted by different groups. 

The application of the empirical correction to the [C/Fe] values based on the evolutionary stage 
of a star and its metallicity is essential for a homogeneous analysis and interpretation. Hence,
in order to have accurate carbon estimates the necessary evolutionary 
correction to the observed carbon abundance values needs to be applied. 

Most of the known CEMP stars  are giants or sub-giants. 
The stars on the red-giant branch (RGB) undergo internal mixing and hence alters 
the surface CN abundances (eg. Aoki et al. 2007; Gratton et al. 2000;  
Placco et al. 2014). 
\citet{placco_et_al_2014_cfe} derived [C/Fe] corrections based on evolutionary stage 
of a star  and its metallicity. 
The application of the empirical correction to the [C/Fe] values from  
\cite{aguado_et_al_2019} resulted in reducing the total number of CEMP stars from 
64 to 12 \citep{Arentsen_et_al_2022}. 
Based on the analysis of a sample of 505 metal-poor ([Fe/H] ${\leq}$$-$2) stars 
with no enhancement of neutron-capture elements from the literature, Placco et 
al. (2014) have shown that the frequency of 
CEMP-no stars ([C/Fe] $\geq$ 0.7) in the Galaxy is increased by $\sim$ 11\% for 
[Fe/H] $\leq$~--3  when proper correction required for the carbon depletion is applied 
to the observed [C/Fe].
A comparison of the  [C/Fe] ratios obtained using corrected and non-corrected 
 carbon abundance  as a function of log \,g is shown in figure 4 for CEMP stars 
from the literature 
 (\citealt{placco_et_al_2014_cfe} and references therein) and our program stars.

Since CEMP-s and CEMP-no stars have very different progenitors, their relative fraction 
provide insight into  the different physical processes at different metallicity.
Hence,  identification of genuine CEMP-no stars and CEMP-s  and  accurate knowledge 
 of  CEMP-s and CEMP-no fractions among  CEMP stars   is crucial.

\begin{figure}
\centering
        \subfigure[]{%
            \label{fig:1a}
            \includegraphics[height=6.5cm,width=7.5cm]{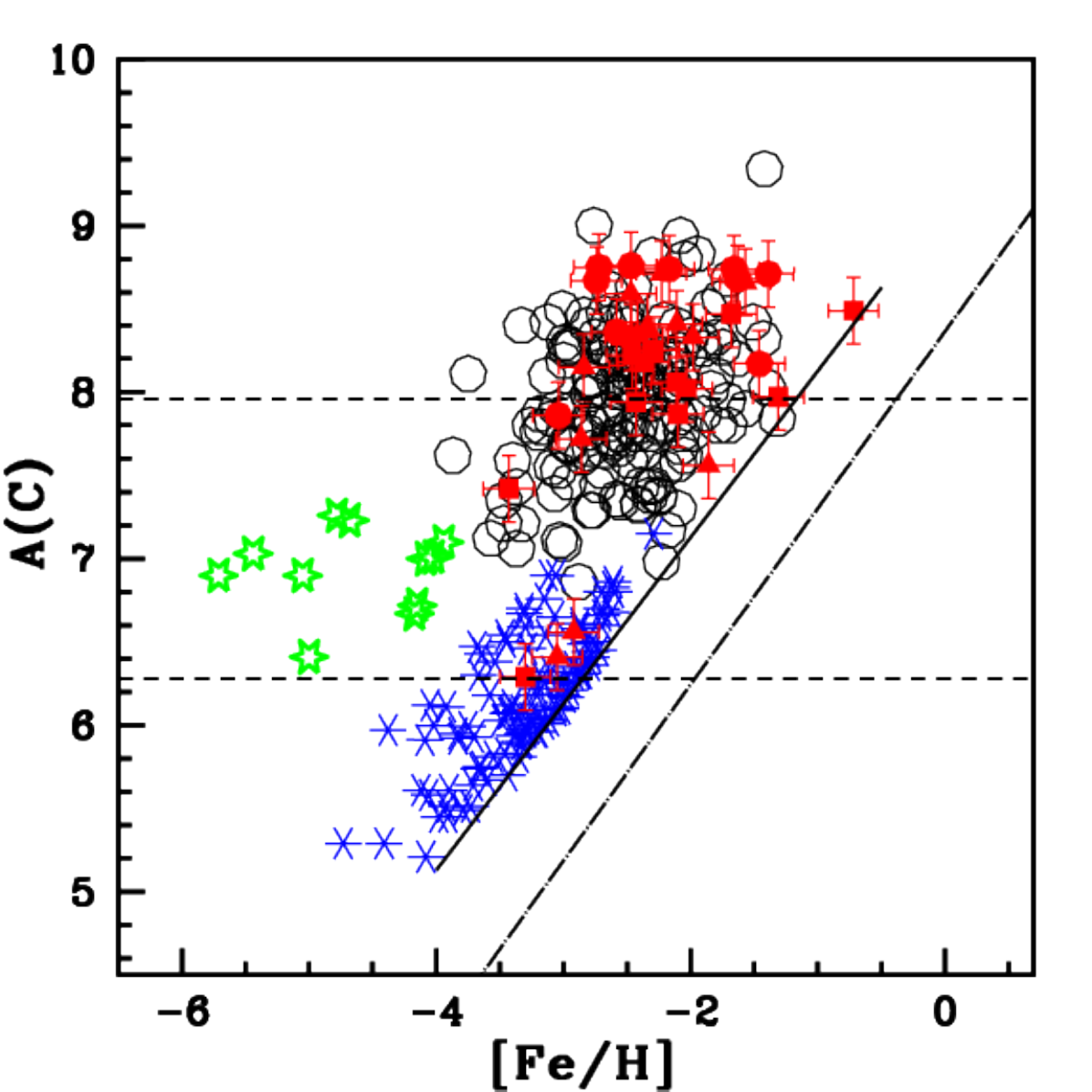}
        }%
        \subfigure[]{%
           \label{fig:1b}
           \includegraphics[height=6.5cm,width=7.5cm]{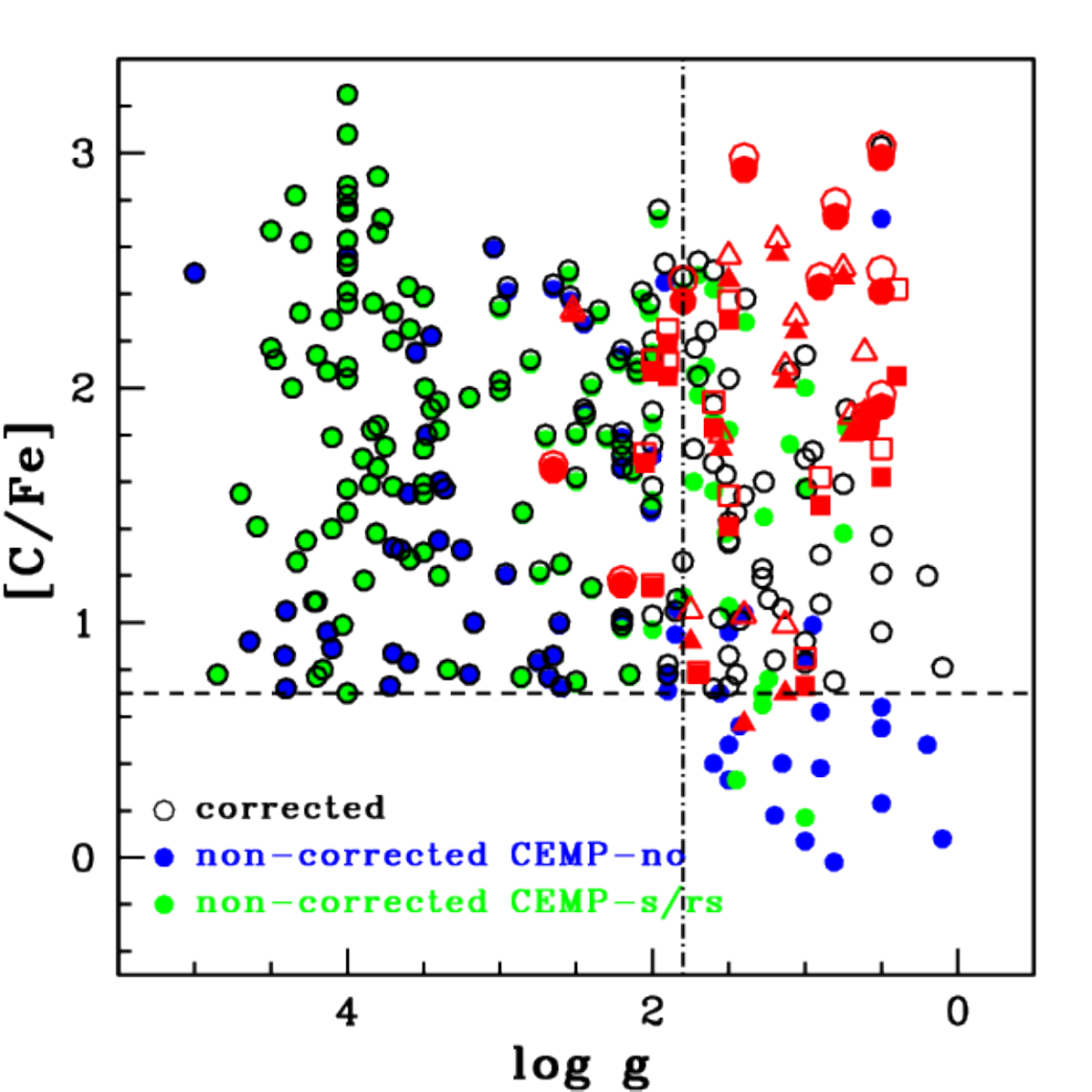}
        }\\ 
\bigskip
\begin{minipage}{12cm}
    \caption{Panel (a):  Absolute carbon abundance A(C) with respect to metallicity  [Fe/H] are 
    plotted for CEMP stars.  Black circles represent  Group I CEMP-s and CEMP-r/s stars, blue 
    crosses Group II CEMP-no stars, and  the green stars represent the Group III CEMP-no stars 
    from the literature (Yoon et al. 2016  and references therein). 
Red triangles represent  CEMP stars from Shejeelammal et al. (2021), Shejeelammal \& Goswami (2021), 
Shejeelammal \& Goswami (2022), and Shejeelammal \& Goswami (2023, in preparation). 
Red rectangles represent  CEMP stars from \citet{Purandardas_et_al_2019}, Purandardas \& Goswami (2021a), 
and Purandardas \& Goswami (2021b). 
The red circles represent  CEMP stars from Goswami et al. (2021), Goswami \& Goswami (2022), and 
Goswami \& Goswami (2023).  The black dashed lines indicate - high and low carbon bands 
at A(C)$\sim$7.96 \& 6.28. The solid line corresponds to  [C/Fe] = 0.70 and the  
long-dash dot line corresponds to   [C/Fe] = 0.
Panel (b): [C/Fe] ratio as a function of log \,g for CEMP stars from the literature 
 (\citealt{placco_et_al_2014_cfe} and references therein). The observed [C/Fe] for   CEMP-no stars  
 are represented by blue filled circles, the CEMP-s and CEMP-r/s stars  by green filled circles, 
 and the corrected  [C/Fe]  by black open circles. Red trinagles indicate CEMP stars from 
 Shejeelammal et al. (2021), Shejeelammal \& Goswami (2021), Shejeelammal \& Goswami (2022) 
and  Shejeelammal \& Goswami (2023, in preparation). Red rectangles represent CEMP stars from Purandardas et al. (2019), Purandardas \& Goswami (2021a) and  Purandardas \& Goswami (2021b). Red circles represent 
CEMP stars from Goswami et al. (2021), Goswami \& Goswami (2022) and Goswami \& Goswami (2023). 
The filled symbols are used for observed values and the open symbols are used for the  corrected 
values. The dashed horizontal line corresponds to  [C/Fe] = 0.7, and the vertical dot-dashed line 
corresponds to log \,g = 1.8. } 
 
\label{fig:6}
\end{minipage}
\end{figure}

 The G-band of CH is commonly used as carbon abundance diagnostics in CEMP stars.  
Most of these estimates are based on LTE assumption. In a recent study,  
\citet{popa_et_al_2023}  have
explored the sensitivity of this feature to NLTE effects considering the  case of
the Sun and red-giant stars. They  found that in the statistical equilibrium of 
the CH molecule this effect is significant, increases with decreasing metallicity, and 
cannot be neglected for precision spectroscopic analysis of C abundances. 
 The C abundance is under-estimated if LTE is assumed. The non-LTE corrections to 
C abundance 
inferred from  the CH band range from +0.04 dex to 0.2 dex for a red giant with 
metallicity [Fe/H] = $-$4.0.

Sample selection is also crucial while  determining CEMP fractions.  
Only stars of similar evolutionary phases should be used to 
compare CEMP samples to ensure similar evolutionary paths,  similar systematic errors 
and to minimize biases.

\section{Concluding Remarks}
\label{sec:conclusion}
Some highlights of the results obtained  from  low- and 
high-resolution spectroscopic analysis of CEMP stars are presented.  
A  new classification scheme put forward by \citet{Goswami_et_al_1_2021} clearly distinguishes 
CEMP-s and CEMP-r/s stars. Detailed abundance  analysis of CH, CEMP-s and CEMP-r/s stars 
re-confirms  pollution from low-mass AGB companion. Parametric model based analysis performed 
for CH, CEMP-s and CEMP-r/s stars also confirm low-mass for their former companion AGB 
stars. Our analysis confirms that a modified  s-process the so-called i-process,  is 
responsible for the observed abundance pattern in CEMP-r/s stars. The i-process models could successfully reproduce the observed abundance in CEMP-r/s stars. 

Peculiar abundances of  HE 1005-1439 present first ever observational evidence of the occurrence 
of dual nucleosynthesis in the companion AGB star, the i-process followed by s-process in various
stages of the AGB evolution, with PIEs triggering i-process followed by s-process during 
third-dredge-up episodes.

Surface chemical composition and spectral signatures of  two extremely metal-poor  CEMP-no stars 
HE~2148$-$2039 and HE~2155$-$2043 reaffirm  multiple origins for CEMP-no stars. 

Knowing the fraction of CEMP stars relative to carbon-normal stars  is important for 
interpretation of populations of metal-poor stars. However, classification and hence the fraction of 
each class is not possible  to estimate accurately  based on low-resolution spectroscopy.
 
The fraction of CEMP stars as a function of metallicity,  how it changes with the different 
  classes of CEMP stars and different stellar  evolutionary phases are some of the outstanding questions that are   yet to find clear answers. 
More homogeneous  analysis of larger  samples are needed
 to unravel the full potential of CEMP stars for Galactic Archaeology.

\begin{acknowledgments}
  AG  acknowledges  the support received from the Belgo-Indian Network for Astronomy \& Astrophysics  project BINA - 2  (DST/INT/Belg/P-02 (India) and BL/11/IN07 (Belgium)). 
\end{acknowledgments}

\begin{furtherinformation}
ORCID Identification of the authors -

0000-0002-8841-8641 (Aruna Goswami)




\end{furtherinformation}

\bibliographystyle{bullsrsl-en}

\bibliography{sample}

\end{document}